\documentclass[aps,prb, twocolumn,superscriptaddress]{revtex4}
\usepackage{graphicx}

\newcommand{\dg}{\dagger}
\newcommand{\Tc}{T_{\text{c}}}
\newcommand{\Mc}{M_{\text{c}}}
\newcommand{\bg}{b^{\dg}}
\newcommand{\la}{\langle}
\newcommand{\ra}{\rangle}
\newcommand{\ua}{\uparrow}
\newcommand{\da}{\downarrow}

\font\elevenmib=cmmib10 scaled 1095
\font\tenmib=cmmib10
\font\eightmib=cmmib10 scaled 800
\font\sixmib=cmmib10 scaled 667
\skewchar\elevenmib='177
\newfam\mibfam

\textfont\mibfam=\tenmib
\scriptfont\mibfam=\eightmib
\scriptscriptfont\mibfam=\sixmib

\def\nd{^{\vphantom{\dagger}}}
\def\yd{^\dagger}
\mathchardef\sigma="711B
\def\ssr#1{{\scriptscriptstyle{\rm #1}}}
\def\vm{{\vec m}}
\def\vh{{\vec h}}

\def\vM{{\vec M}}
\def\nhat{{\hat n}}
\def\ehat{{\hat en}}

\def\ehat{{\hat e}}

\begin{document}

\title{Order and Disorder in AKLT Antiferromagnets in Three Dimensions}

\author{Siddharth A. Parameswaran}
\email{sashok@princeton.edu}
\affiliation{Department of Physics, Joseph Henry Laboratories, Princeton University, Princeton, New Jersey 08544, USA}
\author{S. L. Sondhi}
\email{sondhi@princeton.edu}
\affiliation{Department of Physics, Joseph Henry Laboratories, Princeton University, Princeton, New Jersey 08544, USA}\affiliation{Princeton Center for Theoretical Science, Princeton University, Princeton, New Jersey 08544, USA}
\author{Daniel P. Arovas}
\email{darovas@ucsd.edu}
\affiliation{Department of Physics, University of California at San Diego, La Jolla, California 92093, USA}

\date{July 21, 2008}

\begin{abstract}

The models constructed by Affleck, Kennedy, Lieb, and Tasaki \cite{Affleck87} describe a family of
quantum antiferromagnets on arbitrary lattices, where the local spin $S$ is an integer multiple $M$ 
of half the lattice coordination number.  The equal time quantum correlations in their ground states 
may be computed as finite temperature correlations of a classical $\textsf{O}(3)$ model on the same 
lattice, where the temperature is given by $T=1/M$. In dimensions $d=1$ and $d=2$ this mapping implies
that all AKLT states are quantum disordered. We consider AKLT states in $d=3$ where the nature of the AKLT
states is now a question of detail depending upon the choice of lattice and spin; for sufficiently
large $S$ some form of N{\'e}el order is almost inevitable. On the unfrustrated cubic lattice, 
we find that all AKLT states are ordered while for the unfrustrated diamond
lattice the minimal $S=2$ state is disordered while all other states are ordered. On the frustrated
pyrochlore lattice, we find (conservatively) that several states starting with the minimal $S=3$ 
state are disordered. The disordered AKLT models we report here are a significant addition to
the catalog of magnetic Hamiltonians in $d=3$ with ground states known to lack order on account of
strong quantum fluctuations.

\end{abstract}


\maketitle
\section{Introduction\label{Intro}}
Quantum antiferromagnets have been a fertile field of research for a half century, exhibiting a
great  richness and variety of physical phenomena. In more recent decades, starting with Anderson's
introduction of the RVB state \cite{pwa-rvb} and accelerating with the discovery of the cuprate
superconductors \cite{cuprates-rvb}, much attention has focused on antiferromagnets that allow for 
disordered ground states due to a mix of frustration and quantum fluctuations \cite{review}. 
In an important
step, Affleck \emph{et. al} \cite{Affleck87} showed how to construct models that build in a great deal 
of both these effects by using local projectors - models for which (essentially unique) ground states 
can be determined analytically. These AKLT models have spins given by $S = \frac{z}{2}M$, where $M$ is 
any integer, and $z$ the lattice coordination number. The associated ground states have the added 
feature that their wavefunctions can be written in Jastrow (pair product) form. A general feature of 
such wavefunctions is that the ground-state probability densities can be viewed as Boltzmann weights 
corresponding to a local, indeed nearest neighbor, Hamiltonian for classical spins on the same 
lattice. Using this unusual quantum-classical equivalence we can understand many properties of the 
states via Monte Carlo simulations of the associated classical model.

In $d=1$ and $d=2$, the AKLT states are disordered for \emph{any} spin due to the Hohenberg-Mermin-Wagner
theorem. In particular, the $d=1$ case is the celebrated AKLT chain which realizes the $S=1$ Haldane 
phase. In this paper, we study AKLT states in $d=3$, which are relatively less well-understood than 
their one and two dimensional counterparts. Moreover, in three dimensions, the Hohenberg-Mermin-Wagner 
theorem no longer applies and therefore whether an AKLT state of a given spin is disordered or instead
exhibits long range order is no longer automatic.  Instead a computation is now required to settle this 
question and it is this issue that we address in this paper by a combination of mean-field arguments
and Monte Carlo simulation. Specifically, we discuss the AKLT states on 
the simple cubic and diamond lattices, where there is no (geometrical) frustration, as well on the highly 
frustrated pyrochlore lattice, where  the attendant complications lead to a macroscopic ground state 
degeneracy of the associated classical model. Of course, all the models we study have frustration from
competing interactions.

On the cubic lattice we find that all AKLT states starting with the ``minimal'' (smallest spin)
$S=3$ state are ordered with the standard two sublattice  N{\'e}el pattern. The diamond lattice has
a small coordination number and thus larger fluctuations and we find that on it the
the minimal $S=2$ state is disordered while all higher spin states are ordered with the
two sublattice N{\'e}el pattern. On the pyrochlore lattice, the geometrical frustration of the
lattice plays a significant role. In mean field theory for the companion classical model we find 
a macroscopic number of solutions corresponding to as many energy minima. While the mean field
estimate for the critical spin (transition temperature) already indicates that the minimal $S=3$ 
model on the pyrochlore lattice is disordered, the large number of competing states indicate 
that the true boundary between disorder and some form of order lies at much larger values of spin.
Indeed, a basic simulation leads to a conservative bound in which disordered ground states persist
up to $S=15$. Given the unphysical complexity of the AKLT Hamiltonians at such large spins we do not 
pursue a more precise determination of this boundary in this work. Indeed, readers may take as the
main fruits of our work the identification of the $S=2$ AKLT model on the diamond lattice and the
$S=3$ AKLT model on the pyrochlore lattice as (not too common) instances of three dimensional spin 
Hamiltonians with quantum disordered ground states.

It is worth noting that models with quantum disordered ground states are currently objects of intense
interest in the context of topological order and more specifically, in the context of topological
quantum computing. We note that our disordered models do \emph{not} yield topologically ordered 
states; they do not posess a topological degeneracy or host fractionalized excitations. The 
disordered states herein are described either as fully symmetric valence bond solids or, in the 
long wavelength sense, as quantum paramagnets.  To understand why this is the case, it is instructive 
to recall how a closely related strategy works to produce topologically ordered states in 
$S=1/2$ models, including instances in $d=3$. This strategy,  initiated by Chayes, Chayes and 
Kivelson \cite{Chayes89}, and brought to fruition in work by Raman, Moessner and Sondhi 
\cite{Raman05} works with  spin-$\frac12$ analogs of the AKLT models called Klein models \cite{Klein82}.
Unlike AKLT models, Klein models have many ground states---indeed they select the macroscopically
many nearest neighbor valence bond coverings of a lattice. This selection of a degenerate manifold
underlies the emergence of topological order. More precisely, the work of RMS showed that Klein models
could be controllably perturbed on a family of lattices in order to select a topologically ordered 
(RVB) state in this ground state manifold. In this fashion they could construct $\textsf{SU}(2)$ symmetric 
models with $\textsf{Z}_2$ topological order in $d=2$ but also models with $\textsf{Z}_2$ and $\textsf{U}(1)$ order in 
$d=3$ \cite{fn:xxz-on-pc}.

The rest of this paper is organized as follows: In Section II, we present a brief summary of the AKLT 
construction. We then proceed in Section III to review the mean field analysis of the AKLT states 
\cite{Arovas08}.  We then specialize in Section IV to bipartite lattices, and compute the transition 
temperature for the simple cubic and diamond lattices using Monte Carlo simulations. We determine that 
while the simple cubic lattice exhibits  N{\'e}el order for all choices of $M$ (and thus $S$), the 
diamond lattice allows a quantum disordered state in the $M=1$ ($S=2$) case.  We then go on to discuss 
the AKLT states on the frustrated 3D pyrochlore lattice, and discuss the mean field analysis and classical
ground states in this case. We find that the pyrochlore lattice admits quantum disordered states for many 
values of $M$; while the exact value of $M_{\rm c}$ was not determined, we find evidence from 
simulations that it exceeds $5$, corresponding to $S=15$.

\section{AKLT States: A Brief Review}
The central idea of the AKLT approach \cite{Affleck87}, is to use the idea of quantum singlets to construct correlated quantum-disordered wavefunctions, which are eigenstates of local projection operators. One can then produce many-body Hamiltonians using projectors that extinguish the state, thereby rendering the parent wavefunction an exact ground state, typically with a gap to low-lying excitations. A general member of the family of valence bond solid (AKLT) states can be written compactly in terms of Schwinger bosons \cite{Arovas88}:
\begin{equation}
|\Psi({\cal L}\,;\,M)\ra = \prod_{\la ij\ra} \left(\bg_{i\ua}\bg_{j\da} - \bg_{i\da}\bg_{j\ua}\right)^{\!M} |\,0\,\ra\ .
\end{equation}
This assigns $M$ singlet creation operators to each link $\la ij\ra$ of a lattice $\mathcal{L}$ . The total boson occupancy per site is given by $zM$, where $z$ is the lattice coordination number, and the resultant spin on each site is given by $S = \frac12 zM$. Thus, given any lattice, the above construction defines a family of AKLT states with $S = \frac12 zM$, where $M$ is any integer. The maximum possible spin on any link is then $S^{\text{max}}_{ij} = 2S -M$, and therefore $|\Psi(\mathcal{L};M)\ra$ is extinguished by any Hamiltonian constructed out of projectors $P\nd_J(ij)$ onto link spin $J$, provided $2S-M+1 \le J \le 2S$.
The projectors, which transform as $\textsf{SU}(2)$ singlets, may be written as polynomials in the
Heisenberg coupling ${\vec S}\nd_i\cdot{\vec S}\nd_j$ of order $2S$.  Explicitly, one has
\begin{equation}
P\nd_J(ij)=\prod_{J'=0\atop (J'\ne J)}^{2S} {{\vec S}\nd_i\cdot{\vec S}\nd_j + S(S+1) - 
{1\over 2} J' (J'+1) \over {1\over 2} J(J+1) - {1\over 2} J' (J'+1)}\ .
\end{equation}
 
The AKLT states have a convenient representation in terms of $\textsf{SU}(2)$ coherent states, as first
shown in  Ref.~\onlinecite{Arovas88}.  In terms of the Schwinger bosons, the normalized spin-$S$ coherent state is
given by $| \nhat\ra=(p\,!)^{-1/2}\,(z\nd_\mu b\yd_\mu)^p\,|\,0\,\ra$, where $p=2S$, with
$z=(u\, , \, v)$ a $\textsf{CP}^1$ spinor, with $u=\cos(\theta/2)$ and $v=\sin(\theta/2)\,e^{i\varphi}$. 
The unit vector $\nhat$ is given by $n^a=z\yd \sigma^a z$, where ${\vec\sigma}$
are the Pauli matrices.  In the coherent state representation, the general AKLT state wavefunction is
the pair product $\Psi=\prod_{\la ij \ra} (u\nd_i\,v\nd_j - v\nd_i\,u\nd_j)^M$.  Following Ref.~\onlinecite{Arovas88}, we may write $|\Psi|^2\equiv\exp(-H_{\rm cl}/T)$ as the Boltzmann weight for a classical
$\textsf{O}(3)$ model with Hamiltonian
\begin{equation}
H\nd_{\rm cl}=-\sum_{\la ij \ra} \ln\!\bigg({1-\nhat_i\cdot\nhat_j\over 2}\bigg)\ ,
\label{hcl}
\end{equation}
at temperature $T=1/M$.  All equal time quantum correlations in the state $|\Psi\ra$ may then
be expressed as classical, finite temperature correlations of the Hamiltonian $H\nd_{\rm cl}$.

The consequences of this quantum-to-classical equivalence, which is a general feature of Jastrow
(pair product) wavefunctions, were noted in Ref.~\onlinecite{Arovas88}.  On one and two-dimensional lattices,
the Hohenberg-Mermin-Wagner theorem precludes long-ranged order at any finite value of the discrete
quantum parameter $M$.  Thus, while the $S=2$ Heisenberg model on the square lattice is
rigorously known to have a N{\'e}el ordered ground state \cite{Neves86},
the $S=2$ AKLT Hamiltonian, which includes up to biquartic terms, has a featureless quantum disordered
ground state, called a `quantum paramagnet'. 
In three dimensions, we expect N{\'e}el order for large $M$, corresponding to low
temperatures in the classical model.  If the N{\'e}el temperature for $H\nd_{\rm cl}$ on a given lattice
satisfies $T\nd_{\rm c} > 1$, then $M\nd_{\rm c} < 1$, and all the allowed AKLT states on that lattice
exhibit long-ranged order. 

The issue of whether or not the AKLT states can be in the quantum disordered phase on a given lattice
can be investigated by a combination of mean-field calculations and classical Monte Carlo simulations, which we present below.  

\section{Mean Field Theory}\label{mftsec}
A mean field analysis of the classical model of eqn. \ref{hcl} on bipartite lattices was described in Refs.~\onlinecite{Arovas88,Arovas08}.  In the general case, we may begin with the Hamiltonian of eqn. \ref{hcl},
and we write $\nhat\nd_i=\vm\nd_i+\delta\nhat\nd_i$, with $\langle\nhat\nd_i\rangle=\vm\nd_i$.
Expanding $H\nd_\ssr{cl}$ to order $\delta\nhat\nd_i$, we obtain the mean field Hamiltonian
${\tilde H}^\ssr{MF}=E\nd_0-\sum_i\vh\nd_i\cdot\nhat\nd_i$, where the mean field $\vh\nd_i$ is
given by
\begin{equation}
\vh\nd_i=-{\sum_j}'{\vm_j\over 1-\vm_i\cdot\vm_j}\ ,
\label{hmf}
\end{equation}
where the prime restricts the sum on $j$ to nearest neighbors of site $i$.
Self-consistency then requires
\begin{equation}
\vm\nd_i=\langle\nhat\nd_i\rangle = {\int\!d\nhat\nd_i\>\nhat\nd_i\,
\exp\big(\vh\nd_i\cdot\nhat\nd_i/T\big)\over
\int\!d\nhat\nd_i\>\exp\big(\vh\nd_i\cdot\nhat\nd_i/T\big)}\ ,
\label{mfeqn}
\end{equation}
which yields $\vm\nd_i=m\nd_i\,\vh\nd_i\big/|\vh\nd_i|$, with local magnetization
\begin{equation}
m\nd_i=\coth\bigg({h\nd_i\over T}\bigg)- {T\over h\nd_i}\ .
\label{selfc}
\end{equation}

\section{Unfrustrated Lattices: Simple Cubic and Diamond}
\subsection{Mean Field Transition}
On an unfrustrated, bipartite, lattice a sublattice rotation $\nhat\nd_i\to\eta\nd_i\,\nhat\nd_i$, with $\eta\nd_i=\pm 1$
on the A (B) sublattice, results in a ferromagnetic interaction, and if we posit a uniform local
magnetization $\vm$ we obtain the mean field
\begin{equation}
h={z\, m\over 1+m^2}\ ,
\end{equation}
where $z$ is the lattice coordination number.
This results in a mean field transition temperature $T_{\rm c}^\ssr{MF}=\frac{1}{3}z$, {\it i.e.\/}
$M_{\rm c}^\ssr{MF}= 3 z^{-1}$.  All AKLT states on bipartite lattices in more than two space dimensions
will exhibit two sublattice N{\'e}el order, provided $M>M\nd_{\rm c}$.  According to the mean field
analysis, long ranged order should pertain for $z\ge 3$, which would be satisfied by almost any 
three-dimensional structure.  However, mean field theory ignores fluctuations, hence it overestimates
$T\nd_{\rm c}$ and underestimates $M\nd_{\rm c}$.  Therefore the possibility remains that a quantum
disordered AKLT state may exist in a three-dimensional lattice. We examine two cases, the simple cubic lattice ($z=6$) and the diamond lattice ($z=4$).  We shall address this issue via classical Monte Carlo simulations of our model on both lattices.
We note that, of these, the diamond lattice is the stronger candidate as 
it is more weakly coordinated, and $M_{\rm c}^\ssr{MF}=\frac{3}{4}$ is sufficiently close to threshold that fluctuations are likely to drive the true $M\nd_{\rm c}$ to be greater than unity.

\subsection{Monte-Carlo Simulations}
The classical Hamiltonian $H\nd_{\rm cl}$ of eqn. \ref{hcl} consists of nearest neighbor interactions
$v(\vartheta\nd_{ij})$ where $\vartheta\nd_{ij}=\cos^{-1}(\nhat\nd_i\cdot\nhat\nd_j)$ is the relative angle
between spins on neighboring sites $i$ and $j$, and $v(\vartheta)=-\ln\sin^2\!\big(\frac{1}{2}\vartheta\big)$.
This interaction strongly suppresses ferromagnetic alignment, with a logarithmically infinite barrier,
but has a smooth quadratic minimum $v(\vartheta)\approx\frac{1}{4}(\vartheta-\pi)^2$ when
$\vartheta\approx\pi$.
We have simulated the equivalent ferromagnetic model, with interaction
$v(\vartheta)=-\ln\cos^2\!\big(\frac{1}{2}\vartheta\big)$.

We used a multithread Monte Carlo approach, in which simultaneous simulations with independent initial configurations were used to produce $M$ independent Markov chains each with $N$ configurations \cite{Brown96}, which were then written to a file. For every independent thread, we performed checkerboard sweeps of the lattice using a standard Metropolis Monte Carlo technique \cite{Peczak91}. In each Monte Carlo step, we produced a vector $\delta\nhat$, with length distributed according to a Gaussian and pointing in a random direction,  which was used to generate
a new spin unit vector
\begin{equation}
\nhat'_i= {\nhat\nd_i + \delta\nhat\over |\nhat_i + \delta\nhat|}
\end{equation}
The standard deviation of the Gaussian was adjusted by hand until a significant fraction of proposed moves were accepted (we left it at $\sigma = 0.5$.)

The number of Monte Carlo steps per site (MCS) and the number of independent threads were adjusted to count roughly the same number of autocorrelation times for each sample size\cite{fn2}. For each chain, we obtained the average value of the Binder cumulant, and averaged this across chains to get a single number for each temperature. We estimated the error from the standard deviation of the $M$ \emph{independent} thread averages. This is free of the usual complications of correlated samples inherent in estimating the error from a single chain, and frees us of the need to compute autocorrelation times to weight our error estimate.

Plots were made of the Binder cumulant \cite{Binder81}, defined to be 
\begin{equation}
B = 1 -{\big\langle (\vM^2)^2\big\rangle\over 3\, \big\langle \vM^2\big\rangle^2}\ ,
\end{equation}
where $\vM=\sum_i\nhat\nd_i$ is the total magnetization.

For any system of  Heisenberg spins in the thermodynamic limit, the Binder cumulant has value $\frac{2}{3}$
in the low-temperature (ordered) phase and value $\frac{4}{9}$ in the high-temperature phase. These are easily seen by assuming a gaussian distribution for $|\vM|$ at high temperature, and using the result that all the expectation values of powers of $\vM\cdot\vM$ are equal in the ordered phase. For a finite system, the limiting values continue to be close to these estimates, but the  interpolating behavior is different for each system size; the primary utility from our point of view is that finite-size scaling analysis of $B$ reveals that it has a fixed point at the transition temperature \cite{Binder81}. We may therefore determine $\Tc$ by plotting the Binder cumulant for a series of different lattice sizes, and determining the points where the curves cross.

Before simulating our modified interaction, we checked our code by determining the (known) transition temperatures for the standard Heisenberg model on the simple cubic lattice \cite{Peczak91} and the Ising model on both the diamond and the simple cubic lattices \cite{Fisher67}, as well as comparing the high-temperature susceptibility from simulations to the predictions of the high-temperature expansion \cite{Stanley67}. All these agreed well with the expected values, at least to the accuracy we need to determine whether $\Tc$ is less than or greater than $1$. Recall that if $\Tc<1$, then $\Mc >1$, which means that the minimal AKLT state, with $M=1$, is on the disordered side of the phase transition. 

Using our Monte Carlo simulations, we obtain estimates of $\Tc$ for the families of AKLT states on the simple cubic and diamond lattices. Although our simulation techniques were not particularly sophisticated, they were sufficient to pin down $\Tc$ to a reasonable degree of accuracy, and certainly enough to determine whether $\Tc>1$. Our simulations allow us to estimate that $T_{\rm c}^\ssr{SC} \approx 1.66$ on the simple cubic lattice, and  that $T_{\rm c}^\ssr{D} \approx 0.85$ for the diamond structure.  Therefore, we conclude that while all the simple cubic AKLT states are N\'{e}el-ordered, the minimal ($S=2$) AKLT state in diamond is a featureless quantum disordered state.  

\begin{figure}
\includegraphics[width=8cm]{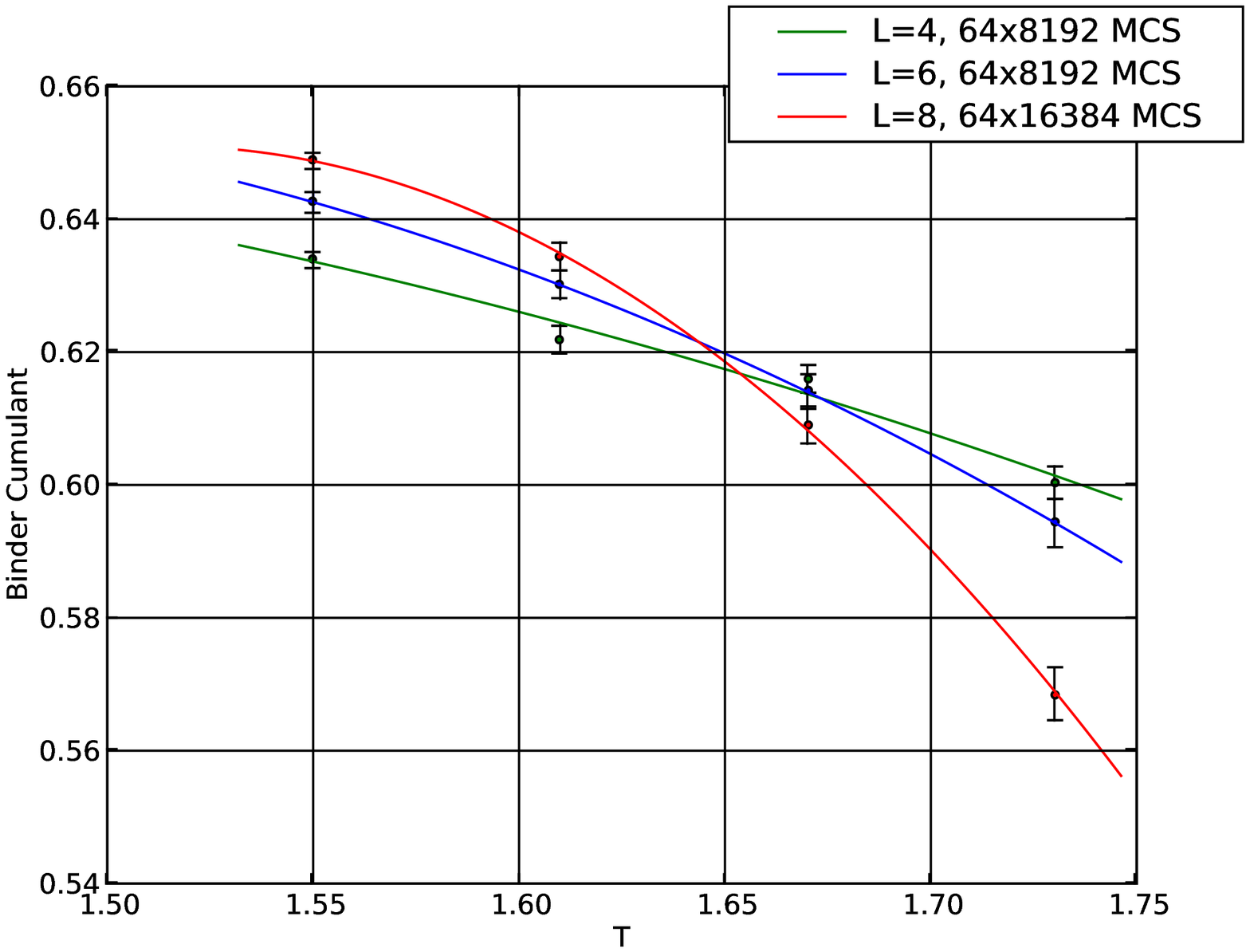}
\includegraphics[width=8cm]{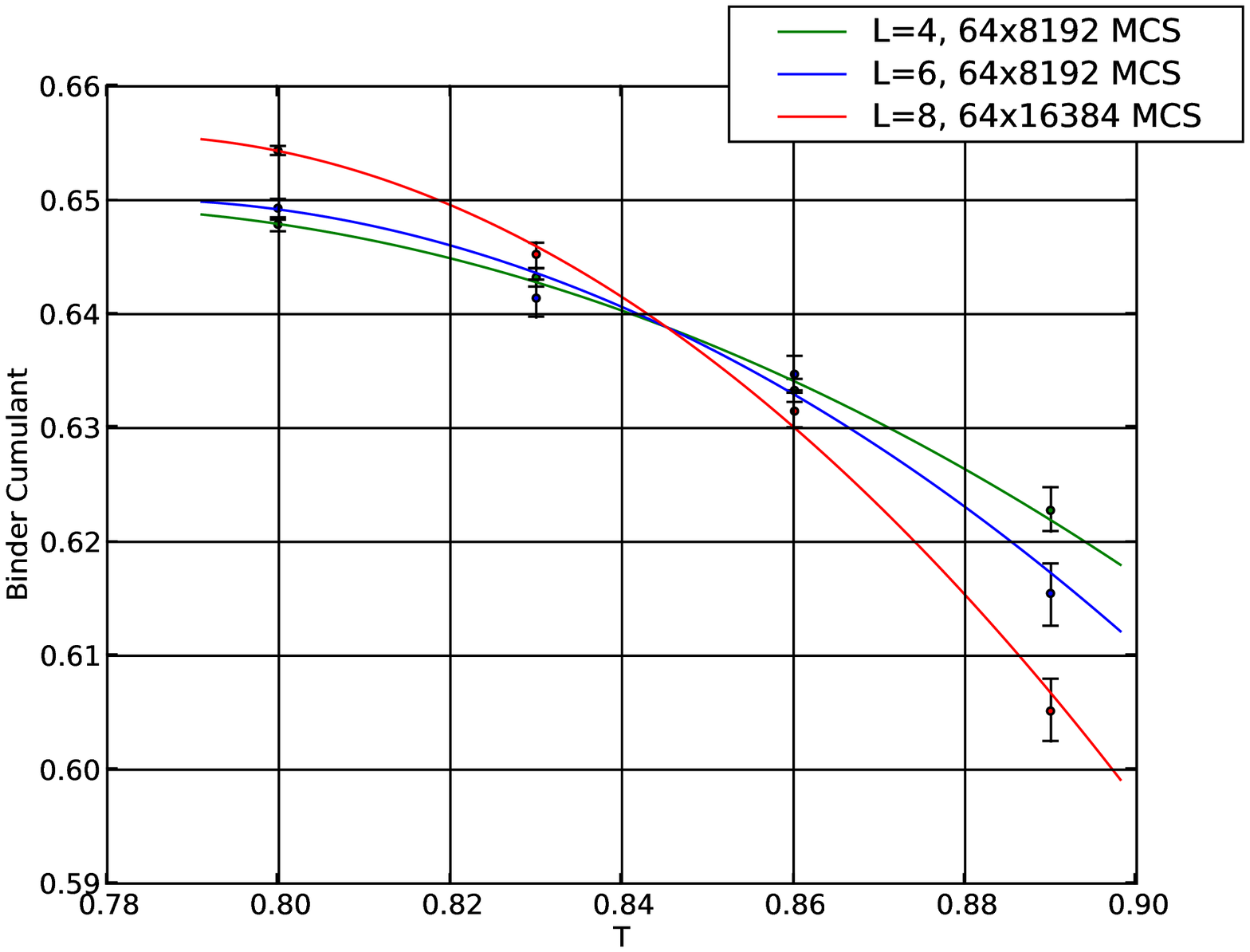}
\caption{Binder cumulant plots for the valence-bond states on the cubic and diamond lattices. The $T$-axis scale is chosen using a rough estimate of $T_{\rm c}$ so as to provide approximately the same window in natural units $T/T_{\rm_c}$ for both cases. In each case, the total number of spins being simulated is $2\cdot L^3$.
We perform a fit of the data (weighted by the error bars) to a parabola  and estimate $T_{\rm c}$ from the intersection of the best-fit lines. We can be reasonably confident that the curves have an intersection from the fact that 
they change order on either side of the crossing, and become separated by more than a standard 
deviation as we move away from the crossing. We obtain $T_{\rm c}\approx 1.66$ for the cubic lattice and $T_{\rm c}\approx0.85$ for 
diamond. }\end{figure}

\section{Frustrated Lattice: The Pyrochlore}

The pyrochlore is a lattice of corner-sharing tetrahedra and can be constructed from the
the diamond lattice by placing a site at the midpoint of each bond, resulting in a quadripartite 
structure. The pyrochlore lattice is highly frustrated from the perspective of of collinear
antiferromagnetism; the canonical nearest-neighbor 
classical Heisenberg antiferromagnet on this lattice has an extensive ground-state degeneracy 
and remains a quantum paramagnet at all temperatures \cite{Moessner98}.

Our problem has a different form for the interaction and hence the results for the nearest
neighbor problem, which build on the high degree of degeneracy for a single tetrahedron,
do not apply. Indeed, as we discuss below, the logarithmic form of the interaction energy leads 
to the selection of a unique single-tetrahedron ground state up to global rotations.
However, the full lattice still exhibits a substantial ground state degeneracy on account of
its open architecture indicating an anomalously low transition temperature which we roughly
bound from above by $T \approx 0.2$.
We now turn to the details of these assertions.

\begin{figure}
\includegraphics[width=7cm]{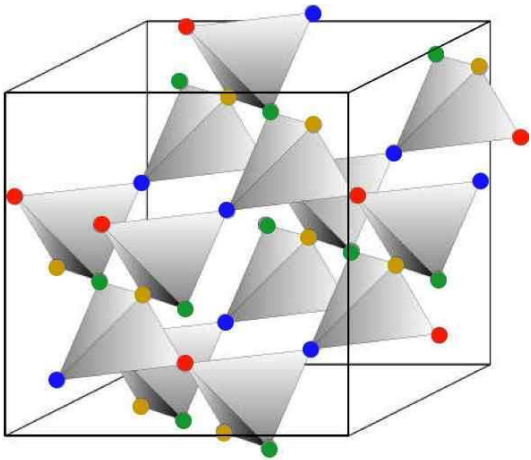}
\caption{ The quadripartite pyrochlore lattice, which is formed out of corner-sharing tetrahedra.}
\end{figure}

\subsection{Single-Tetrahedron Ground States}
Numerical minimization on a single tetrahedron finds the lowest-energy configuration to be one where each pair of spins make an angle $\vartheta_{ij} = \cos^{-1}(\nhat\nd_i\cdot\nhat\nd_j) =\cos^{-1}\left(-\frac13\right)$. This means that the spins are pointing either towards or away from the corners of a regular tetrahedron in three-dimensional spin space. We  proceed to search for soft modes, by expanding the energy to quadratic order and studying the resulting normal modes. We find that there is a pair of soft modes corresponding to global rotations, and another where three spins rotate about the axis defined by the fourth. The latter mode leads to a degeneracy of ground states of the full lattice as discussed below.

We note that for the Heisenberg antiferromagnet with interaction $\nhat_i\cdot\nhat_j$, the single tetrahedron Hamiltonian is
$H_\Gamma=\big({\vec M}_\Gamma\big)^2$, where ${\vec M}_\Gamma=\sum_{i\in\Gamma}\nhat_i$ is a sum of the spin vectors over all sites in the tetrahedron $\Gamma$.   The ground state manifold ${\vec M}_\Gamma=0$ is then five-dimensional, since one can choose any two vectors $\nhat_A$ and $\nhat_B$, then take $\nhat_C=-\nhat_A$ and $\nhat_D=-\nhat_B$.  The four freedoms associated with choosing $\nhat_A$ and $\nhat_B$ are then augmented by another freedom to rotate the $C$ and $D$ spins about the direction $\nhat_A+\nhat_B$.   A large-$N$ analysis\cite{Isakov04} finds that the $\textsf{O}(N)$ pyrochlore
antiferromagnet is paramagnetic down to $T=0$.

\subsection{Ground States on the Full Lattice}
There are many ways in which we can construct degenerate states on the lattice that simultaneously satisfy the
minimum-energy constraint on every tetrahedron.  We begin by describing the simplest
such states which form a discrete family. To this end, label the four spins defined by the 
single-tetrahedron constraint (with a fixed joint orientation) as $A$, $B$, $C$ and $D$. If we use only
these four orientations for each spin, we have the constraint that none of them can occur twice
on the same tetrahedron; this translates to the statement that spins on neighboring links must 
be different. This is the same constraint as for ground states of the antiferromagnetic 4-state 
Potts model. We therefore conclude that one family of ground states of the classical Hamiltonian 
on the pyrochlore lattice  are in a one-to-one correspondence with the ground states of the 
4-state Potts antiferromagnet on the pyrochlore lattice. Readers familiar with the lore on the
kagom\'{e} problem\cite{Huse92} will recognize the resemblance to the planar ground states there which are in
correspondence to ground states of the 3-state Potts model. As in the kagom\'{e} problem, from this 
set of ground states others can be constructed by identifying sets of spins which can be locally 
rotated by an arbitrary amount at zero energy cost. These are sets of spins, say of type $B$, $C$
and $D$ which are connected to other spins solely by spins of type $A$. Clearly one can rotate this
set by an angle about the $A$ axis at zero energy cost.

While we have not parametrized the full, continuous, space of ground states an extensive lower
bound on the degeneracy of the `Potts submanifold' of ground states can be obtained as follows. 
First, we note that the number of allowed configurations of the 3-state Potts model on a 
kagom\'{e} lattice with $M$ sites is given \cite{Huse92, Baxter70} by
$g_{\text{k}}\approx(1.20872)^{(2M/3)}$.  Next we partition the pyrochlore into four sublattices, so that the sites that lie on a single tetrahedron are each on different sublattices.  Choose one sublattice, and fix the spins on that sublattice to be one of the four types (say A.)  Now, looking down through the tetrahedra, one sees alternating layers of triangular and kagom\'{e} planes; the kagom\'{e} planes are made up of B, C, and D spins, while the triangular planes are made up of A spins. In each kagom\'{e} plane, we have $M$ spins, whose configurations are those of the $3$-state Potts model. If we now let $N_{\text{k}}$ be the number of kagom\'{e} planes, we must have that $M\cdot N_{\text{k}} = \frac34 N$, where $N$ is the total number of spins in the system. We then have for the number of states in this restricted submanifold
\begin{equation}
	g_{\text{restricted}} = 4\cdot g_{\text{k}}^{N_{\text{k}}} \approx 4\cdot(1.20872)^{N/2}
\end{equation}
where the factor of $4$ stems from the fact that we can choose any one of the four spins to be fixed in the triangular planes. Since we've restricted ourselves to considering a certain submanifold of the ground states in the above argument, it is clear that we have obtained a lower bound for the degeneracy of the Potts submanifold.

\subsection{Bounds on $T_c$}

Each of the ground states identified above can serve as a basis for a mean-field treatment
of the system and all of them yield the same $T_{\rm c}^\ssr{MF}$. This vast set of ``soft modes''
is, of course, a signature that the true $T_c \ll T_{\rm c}^\ssr{MF}$. Thus we may begin with a 
calculation of $T_{\rm c}^\ssr{MF}$ which can serve as an upper bound on the true $T_c$. 

Consider a spin at site $i$ in the pyrochlore lattice. Expanding in small fluctuations about any ground state, we have the same mean-field Hamiltonian as in the general mean field {\it Ansatz\/} of section \ref{mftsec}, with the mean field given by eqn. \ref{hmf}. In a mean-field treatment, each of the neighbor spins $\nhat_j$ is to be replaced by its average $\vm_j = \langle\nhat_j\rangle = m\ehat_j$ in the particular ground state that we are considering. In any ground state, we note that the angle between any pair of nearest neighbors is $\vartheta_{ij} = \cos^{-1}\left(-\frac13\right)$. In addition, the spins on a tetrahedron add to zero, which allows us to write ${\sum_j}'\vm_j = -\vm_i$. If we further recall that each spin lies on exactly two tetrahedra, we obtain the following expression for the mean field acting at site $i$:
\begin{equation}
\vh\nd_i=-{\sum_j}'{\vm_j\over 1-\vm_i\cdot\vm_j} =\frac{ 2 m}{1 +\frac{m^2}{3}}\ehat_i\label{MFPC}
\end{equation}
Note that we have only made use of the \emph{local} structure of the ground state, and so our treatment here is relevant for the transition into any state in the ground state manifold. Substituting the mean field in eq. \ref{MFPC} into the self-consistency condition  (eqn.\ref{selfc}), we find, in a manner similar to the bipartite case, that the mean-field estimate of the transition temperature is
$T_{\rm c}^\ssr{MF} = \frac{2}{3}$. From this alone we conclude that the $M=1$ state on the 
pyrochlore lattice is quantum disordered. 

There is little question that the actual $T_c$ is much lower than the mean-field estimate
and therefore $M_{\rm c}$ is much higher than $\frac32$, allowing many more quantum disordered
states. As is familiar from other highly frustrated magnets, where the ground-state manifold 
encompasses a vastly degenerate set of states the transition will be driven by the 
`order-by-disorder' mechanism wherein a particular state or subset of states is favored by 
entropic effects at low temperatures. This is a weak effect and hence $T_c$ is typically
a small fraction of $T_{\rm c}^\ssr{MF}$. 
From coarse Monte Carlo simulations we find evidence that $T_c < 0.2$ corresponding to 
$S_c > \frac{1}{2} zM = 15$. This leads already to an AKLT Hamiltonian that is a 
degree-$60$ polynomial in the spins and thus there is little reason in the current context
to locate the transition or the nature of the ordered phase with greater precision.

However, we note that the same set of ground states arises in the the classical Heisenberg 
model with nearest neighbor bilinear and biquadratic interactions with the latter chosen to
disfavor collinearity. This is a physically plausible model and we will report a fuller
investigation of it elsewhere \cite{wip}.

\section{Concluding Remarks}
To summarize, we have studied AKLT states on two unfrustrated and one frustrated lattice
in $d=3$ by a combination of mean-field theory and Monte Carlo simulations for the associated
classical models. We find that the simple cubic lattice is  N{\'e}el ordered at all values of 
the singlet parameter $M$ and spin $S$; the diamond lattice, on the other hand, is 
quantum-disordered for $M=1$ ($S=2$), and  N{\'e}el ordered for $M>1$. On the pyrochlore lattice
we find that the $M=1$ ($S=3$) model is definitely disordered and the boundary between
disorder and order very likely lies above $M=5$.

While quantum-disordered ground states in low (i.e. one and two dimensions) have often been
discussed, three dimensions has historically been the province of long range order. Hence
our disordered models on the diamond and pyrochlore lattices significantly expand the set of 
possibilities for quantum ground states of models with Heisenberg symmetry in $d=3$.

In recent work, one of us has generalized the AKLT construction to $\textsf{SU}(N)$ spins 
\cite{Arovas08}. In the near future, we intend to investigate the $\textsf{SU}(4)$ simplex 
state on the pyrochlore lattice introduced in this work by methods similar to the ones
used in the present paper \cite{wip2}.

\begin{acknowledgments}
It is a pleasure to thank Fiona Burnell and Chris Laumann for many discussions and insightful suggestions.  David Huse provided invaluable guidance on writing efficient Monte Carlo code. We are also grateful to Roderich Moessner for several conversations about the structure of ground states and the potential for order-by-disorder on the pyrochlore lattice. SAP acknowledges the hospitality of the Institute for Mathematical Sciences, Chennai, India, and the Ecole de Physique des Houches, Les Houches, France, where parts of this work were completed.  This work was supported in part by NSF Grant No. DMR 0213706 (SLS).
\end{acknowledgments}


\begin{thebibliography}{10}

\bibitem{Affleck87}
I. Affleck, T. Kennedy, E. H. Lieb, and H. Tasaki, {\sl Phys. Rev. Lett.} {\bf 59}, 799 (1987);
{\sl Comm. Math. Phys} \textbf{115}, 477 (1988).

\bibitem{pwa-rvb}
P. W. Anderson, {\sl Mater. Res. Bull.} {\bf8}, 153 (1973); P. Fazekas and P. W. Anderson, {\sl Phil. Mag.} \textbf {30}, 423 (1974).

\bibitem{cuprates-rvb}
More precisely the ideas of P. W. Anderson,  {\sl Science}
\textbf{235}, 1196 (1987).

\bibitem{review}
For a review, see S. Sachdev, in {\sl Low dimensional quantum field theories for condensed matter physicists}, Yu Lu, S. Lundqvist, and G. Morandi eds., World Scientific, Singapore (1995); cond-mat/9303014.
 
\bibitem{Chayes89}
J.T. Chayes, L. Chayes, and S.A. Kivelson, {\sl Commun. Math. Phys.} {\bf 123}, 53 (1989).

\bibitem{Raman05}
K.S. Raman, R. Moessner, and S.L. Sondhi, {\sl Phys. Rev. B} {\bf 72}, 066413 (2005).

\bibitem{Klein82}
D.J. Klein, {\sl J. Phys. A} {\bf 15}, 661 (1982).

\bibitem{fn:xxz-on-pc}
See also the closely related work on an XXZ model in $d=3$. M. Hermele, M.P.A. Fisher, and L. Balents, {\sl Phys. Rev. B} {\bf 69}, 064404 (2004).

\bibitem{Arovas08}
D. P. Arovas, {\sl Phys. Rev. B} {\bf 77}, 104404 (2008).

\bibitem{Arovas88}
D. P. Arovas, A. Auerbach, and F. D. M. Haldane, {\sl Phys. Rev. Lett.} {\bf 60}, 531 (1988).

\bibitem{Neves86}
E. J. Neves and J. F. Perez, {\sl Phys. Lett.} {\bf 114A}, 331 (1986).

\bibitem{Brown96}
This approach was suggested to us by D. Huse; it is also described in Robert G. Brown and Mikael Ciftan , {\sl Phys. Rev. B} {\bf 54}, 15860 (1996).


\bibitem{Peczak91}
P. Peczak, A.M. Ferrenberg and D.P. Landau, {\sl Phys. Rev. B} {\bf 43}, 6087 (1991).


\bibitem{fn2}
System configurations were recorded after each lattice sweep, so 1 MCS is the natural unit of time along the Markov chains. A precise determination of the autocorrelation time was not performed, but plots of the error estimate were made for blocks of increasing length and initial position along the chain, which allowed us to check the convergence of physical quantities; the final block, consisting of the latter half of the chain, was used to perform averages in each thread. 


\bibitem{Binder81}
K. Binder, {\sl Z. Physik B} {\bf 43}, 119 (1981).



\bibitem{Fisher67}
M.E. Fisher, {\sl Rep. Prog. Phys.} {\bf 30}, 615-730 (1967).  

\bibitem{Stanley67}
H.E. Stanley, {\sl Phys. Rev.} {\bf 158}, 546 (1967).

\bibitem{Moessner98}
R. Moessner and J.T. Chalker, {\sl Phys. Rev. Lett.} {\bf 80}, 2929 (1998);  {\sl Phys. Rev. B} {\bf 58}, 12049 (1998).

\bibitem{Isakov04}
S. V. Isakov, K. Gregor, R. Moessner, and S. L. Sondhi, {\sl Phys. Rev. Lett.} {\bf 93}, 167204 (2004).

\bibitem{Huse92}
D. A. Huse and A.D.  Rutenberg, {\sl Phys. Rev. B} {\bf 45}, 7536 (1992)

\bibitem{Baxter70}
R.J. Baxter, {\sl J. Math. Phys} {\bf 11}, 784 (1970.) The problem discussed here is the three-coloring problem on the 2d hexagonal lattice, which is equivalent to the model  discussed in Ref. ~\onlinecite{Huse92}.

\bibitem{wip}
S. A. Parameswaran, S.L. Sondhi, D.P. Arovas, R. Moessner, work in progress.

\bibitem{wip2}
S. A. Parameswaran, S.L. Sondhi, D.P. Arovas, work in progress.

\end{thebibliography}

\end{document}